# CP-violation effects in neutral meson oscillations in the left-right weak interaction model

A.P. Serebrov, O.M. Zherebtsov, A.K. Fomin, R.M. Samoilov, N.S. Budanov

*National Research Center "Kurchatov Institute" - Petersburg Institute of Nuclear Physics, 188300, Gatchina, Russia*

**e-mail: serebrov_ap@pnpi.nrcki.ru*

## Abstract

An analysis of the latest, most accurate experimental data on neutron decay indicates the need to expand the Standard Model by introducing an admixture of the right vector boson $W_R$ with a mixing angle of $\zeta = -0.039 \pm 0.014$ with the left vector boson $W_L$ and a ratio of the squares of the masses of $W_1$ and $W_2$ equal to $\delta = 0.070 \pm 0.010$. In this regard, the possibility of describing CP-violation effects in neutral meson oscillations within the framework of the left-right weak interaction model with parameters $\delta$ and $\zeta$ was investigated. It was shown that within this model, CP violation effects in the decays of $K^0$-mesons, $D^0$-mesons, $B^0$-mesons, and $B_s^0$-mesons can be successfully described. The results of calculations within the extended left-right model with parameters $\delta$ and $\zeta$ are confirmed by experimental results. Thus, the nature of CP violation is related to the presence of a right-handed vector boson admixture.

## 1. Introduction

Precision studies of neutron decay enable the search for deviations from the Standard Model. Neutron decay has been studied for over half a century. The measurement accuracy has steadily increased and currently stands at $4 \cdot 10^{-4}$ for the neutron lifetime and $10^{-3}$ for the decay asymmetry. This research process involves numerous groups at renowned research centers around the world. Research in Russia has made a significant contribution since the middle of the last century, particularly through ultracold neutron research and precise measurements of the neutron lifetime, as well as neutrino asymmetry measurements.

A brief overview of the current state of neutron decay research is presented in [1], where it is shown that there is a deviation from the Standard Model at the level of 3.7σ. At the same time, this work shows that the results of neutron lifetime and decay asymmetries measurements can be successfully reconciled within the framework of the simplest left-right manifesto model of weak interaction [2-4]. From this analysis in [1], the model parameters were obtained: the mixing angle and the ratio of the squares of the masses of the left and the mixed in right vector bosons. In the same work, studies of proton decay in nuclei, the so-called superallowed
$0^+$-$0^+$ transitions, which have been carefully studied for more than 30 years [5,6], are discussed. These studies make it possible to independently determine the element of the CKM matrix. Moreover, it was shown that the value obtained from these studies differs from the value obtained from the unitarity of the first row of the CKM matrix by 2.4σ, with $V_{ud}^{unit} = \sqrt{1 - V_{us}^2 - V_{ub}^2}$. Although this difference does not exceed 3σ it is alarming, as the question of the unitarity of the CKM matrix is fundamentally important, especially in conjunction with the deviation from the SM in neutron decay. An attempt to reconcile the value of $V_{ud}^{n}$ obtained from neutron decay and $V_{ud}^{00}$ from proton decay in nuclei within the framework of the simplest left-right manifest model failed. However, it turned out that better agreement could be obtained by reversing the sign of the mixing angle when the vector boson charge was reversed. This led to the so-called extended version of the left-right model, where the mixing scheme has the following form:

$$\begin{pmatrix} W_L^{\pm} \\ W_R^{\pm} \end{pmatrix} = \begin{pmatrix} \cos\zeta & \mp\sin\zeta \\ \pm\sin\zeta & \cos\zeta \end{pmatrix} \begin{pmatrix} W_1^{\pm} \\ W_2^{\pm} \end{pmatrix} \quad (1.1),$$

where ζ is the mixing angle of the flavor states $W_L$ and $W_R$, and δ is the ratio of the squared masses of the states $W_1$ and $W_2$.

In the extended left-right model, the following parameter values were obtained from precision studies of neutron decay: $\delta = 0.070 \pm 0.010$, $\zeta = -0.039 \pm 0.014$. From the ratio of the squared masses of the states $W_1$ and $W_2$, it follows that the mass $M_{W_R} = 304^{+24}_{-20}$ GeV. However, in work [1] it was shown that collider experiments do not contradict the results of this analysis, since there is a mixed state of the right-handed vector boson and the left-handed vector boson, and the resonance should be suppressed by more than two orders of magnitude and therefore was not detected.

In the left- and right-handed vector boson mixing scheme, the plus sign is chosen for the particle ( $W^-$), and the minus sign for the antiparticle ($W^+$) at the sine in the top row. It should be noted that the scheme we are considering has an important difference compared to the commonly used scheme, which does not take into account the different mixing signs for particles and antiparticles. In the new scheme, we essentially introduce a difference in the interaction of quarks with $W^-$ and $W^+$, i.e. for particles and antiparticles due to the different signs of $\zeta$, which will lead to CP violation.

It should be clarified that the discrepancy between the values of $V_{ud}^{00LR}$ and $V_{ud}^{nLR}$ can be interpreted as a violation of CP-invariance, since $V_{ud}^{nLR}$ corresponds to the transition of the d-quark to the u-quark, and $V_{ud}^{00LR}$ corresponds to the transition of the u-quark to the d-quark (Fig. 1). Note that the difference is that the neutron decay occurs via a negative vector boson in the mixed state $W_1^-(W_2^-)$, while the proton decays in the nucleus via a positive vector boson in the mixed state $W_1^+(W_2^+)$, and the sign of the mixing angle is opposite. The asymmetry value is

$$A_{p-n}=\frac{(V_{ud}^{00LR})^2-(V_{ud}^{nLR})^2}{(V_{ud}^{00LR})^2+(V_{ud}^{nLR})^2}=(-3.2\pm1.6)\cdot10^{-3} \qquad (1.2)$$

Unfortunately, the accuracy of this parameter's determination is still insufficient, but its value is of the same order of magnitude as the CP-violation parameters in K-meson decays. Therefore, it is advisable to analyze CP-violation processes in K-meson decays within the framework of the extended left-right model, using the δ and ζ parameters from neutron decay. The next section of this paper is devoted to this issue.

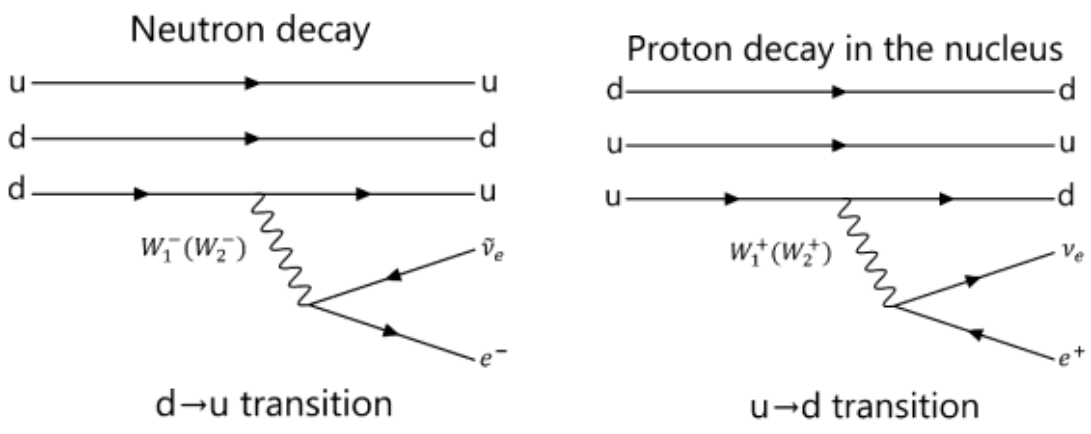

Fig. 1. The process of neutron decay and the process of proton decay in the nucleus. The direct transition from the d-quark to the u-quark and the reverse transition from the u-quark to the down-quark.

## 2. Analysis of CP-violation processes in neutral meson oscillations within the framework of the extended left-right weak interaction model

First, we estimate the integral of the CP-violation effect for the system $K^0\ \bar{K}^0$. During the oscillations of the system $K^0\ \bar{K}^0$, decay into the state $e^-\pi^+\bar{\nu}$ or into the state $e^+\pi^-\nu$ can occur.

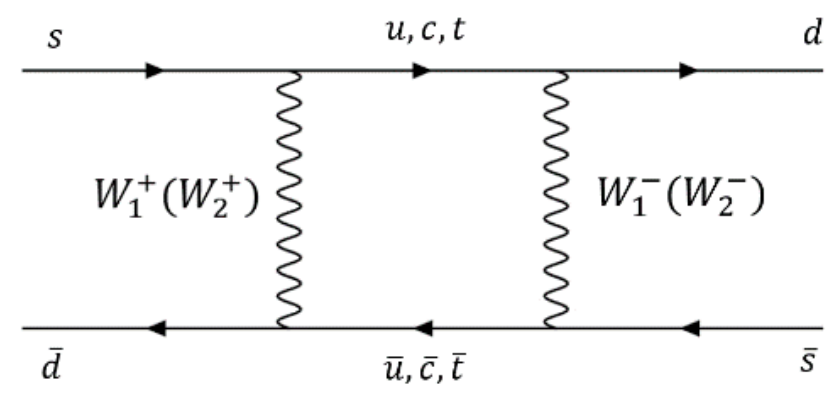

The Hamiltonian of the weak interaction can be represented in the same general form as for $0^+ \leftrightarrow 0^+$ transitions. However, K-mesons are pseudoscalar particles with spin zero and negative parity; these are $0^- \leftrightarrow 0^-$ transitions. Therefore, the sign of $\zeta$ is reversed compared to $0^+ \leftrightarrow 0^+$ transitions. For kaon-antikaon transitions, we must write the Lagrangian with vector current:

$$H_V^N=\bar{e}\gamma_\mu\left(C_V+C'_V\gamma_5\right)\nu\cdot\bar{\pi}\gamma_\mu K^0 \quad (2.1)$$

where the decay $W_1^+(W_2^+)$ is associated with the ratio

$$\left|C_V\right|^2+\left|C'_V\right|^2=G_F^2\left|V_{us}\right|^2\left(1+\left(\delta-\zeta\right)^2\right) \quad (2.2),$$

and the decay $W_1^-(W_2^-)$ is associated with the ratio

$$\left|C_V\right|^2+\left|C'_V\right|^2=G_F^2\left|V_{us}\right|^2\left(1+\left(\delta+\zeta\right)^2\right) \quad (2.3).$$

When decaying into the final state $e^+\pi^-\nu$, we have, up to quadratic terms,

$$\Gamma^{W^+}\propto\left|V_{us}\right|^2\left|f_+\right|^2\left[1+\left(\delta-\zeta\right)^2\right] \quad (2.4)$$

When decaying into the final state $e^-\pi^+\bar{\nu}$, we have, up to quadratic terms,

$$\Gamma^{W^-}\propto\left|V_{us}\right|^2\left|f_+\right|^2\left[1+\left(\delta+\zeta\right)^2\right] \quad (2.5)$$

Thus, we obtain the formula for T-violating asymmetry:

$$A_T=\frac{\Gamma\left(\bar{K}^0\to e^+\pi^-\nu\right)-\Gamma\left(K^0\to e^-\pi^+\bar{\nu}\right)}{\Gamma\left(\bar{K}^0\to e^+\pi^-\nu\right)+\Gamma\left(K^0\to e^-\pi^+\bar{\nu}\right)} \quad (2.6)$$

which is equal to:

$$A_T=\frac{1+\left(\delta-\zeta\right)^2-\left(1+\left(\delta+\zeta\right)^2\right)}{2\left(1+\delta^2+\zeta^2\right)}\approx-2\delta\zeta \quad (2.7)$$

Using the values obtained earlier, we obtain the value for asymmetry

$$A_T^{LR}=-2\delta\zeta=(5.5\pm2.1)\times10^{-3}(2.6\sigma) \qquad (2.8)$$

This value is within the available accuracy and agrees with the experimentally measured asymmetry [7].

$$A_T^{\exp}=\left(6.6\pm1.3\pm1.0\right)\times10^{-3}(4\sigma)\,. \qquad (2.9)$$

Now let's take a closer look at the CP violation process in neutral meson oscillations.

The process of neutral meson oscillations involves the transition of particles to antiparticles and back from antiparticles to particles. Transitions between neutral particles do not violate CPT invariance, but CP violation is possible during these transitions.

CP- violation with mixing of right-handed and left-handed W bosons is possible for neutral mesons such as $K^0-\bar{K}^0$, $D^0-\bar{D}^0$, $B^0-\bar{B}^0$, $B_s^0-\bar{B}_s^0$ mesons. This also applies to $n-\bar{n}$ oscillations. The specific structure of transitions between quark states is presented below:

$$K^0(d\bar{s})-\bar{K}^0(\bar{d}s),\ D^0(c\bar{u})-\bar{D}^0(\bar{c}u),\quad B^0(d\bar{b})-\bar{B}^0(\bar{d}b),\ B_s^0(s\bar{b})-\bar{B}_s^0(\bar{s}b),\ n(udd)-\bar{n}(\bar{u}\bar{d}\bar{d}) \quad (2.10)$$

There are experimental constraints on the violation of CPT-invariance. These can be divided into direct constraints from experiments with $\pi^+$ and $\pi^-$-mesons and indirect constraints from experiments with $K^0-\bar{K}^0$ oscillations. In experiments with $\pi^+$ and $\pi^-$ mesons, the lifetimes of the particle and antiparticle can be measured directly and compared.

CPT violation in $\pi^+$ and $\pi^-$ - decays [8] has already been experimentally limited to a precision level of $10^{-3}$.

$$\frac{\left(\tau_{\pi^+}-\tau_{\pi^-}\right)}{\tau_{average}}=(7.1\pm5.5)\times10^{-4} \quad (2.11)$$

The same restrictions proving CPT invariance exist from the decays $K^0,\bar{K}^0$ [9] where

$$\frac{\left(\Gamma_{K^0}-\Gamma_{\bar{K}^0}\right)}{\Gamma_{\text{average}}}=\left(5.4\pm5.4\right)\times10^{-4} \quad (2.12)$$

However, this is an indirect limitation, since it is not possible to measure the lifetimes separately for $K^0$ and $\bar{K}^0$ because $K^0$ and $\bar{K}^0$ are in a regime of interconversion, and the lifetime of each of them cannot be determined, as demonstrated by the experimental results in Fig. 2.

Estimation (2.12) was obtained in [9], where a mixing matrix with different decay probabilities for $K^0$ and $\bar{K}^0$ on the matrix diagonal was considered to account for CPT violation, but CP violation was taken into account in off-diagonal matrix elements. This result (2.3) should be interpreted as follows: after taking into account CP-violation in off-diagonal matrix elements within the CM framework, nothing remains for CPT violation in diagonal matrix elements with the specified accuracy. Therefore, one should rely on this experimental result and adopt the position, based on the results of the experiment [9], that CPT invariance is preserved. Thus, in the extended left-right model, CP invariance is violated and CPT invariance is conserved.

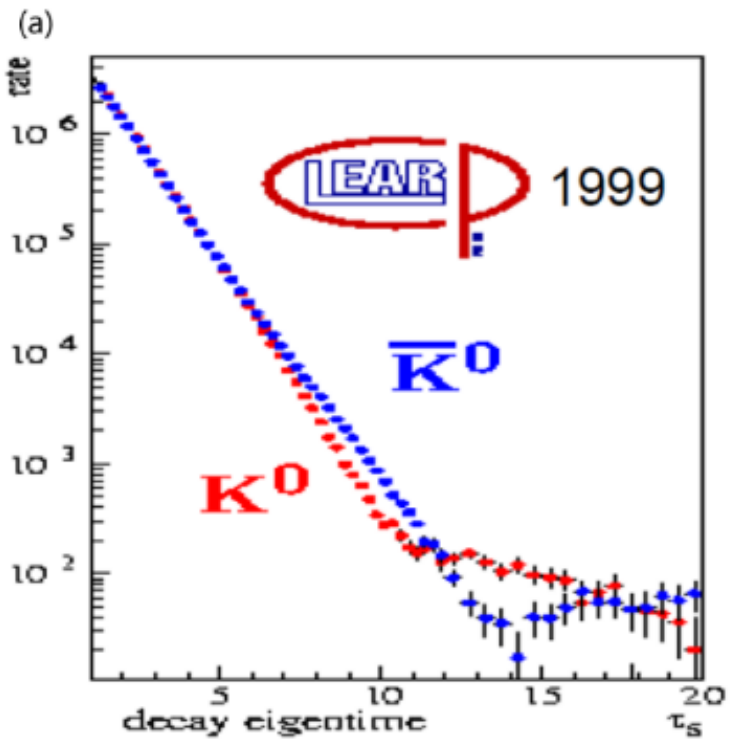

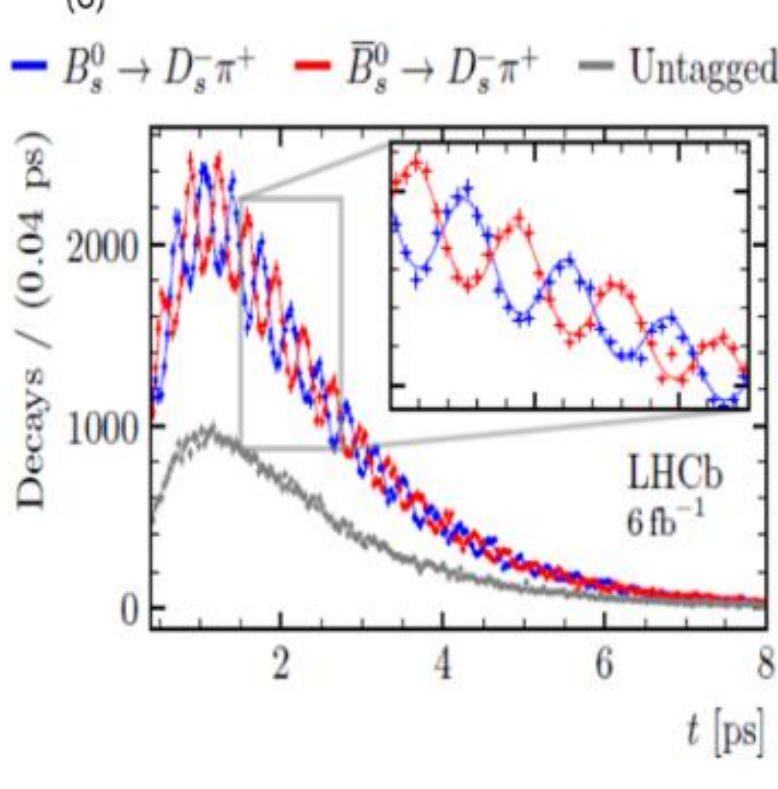

Fig. 2. Experimental results of flavor oscillations: a) for $K^0$meson oscillations [10], b) for $B_s^0\bar{B}_s^0$oscillations. [11]

The constraints on the model parameters δ and ζ for charged and neutral particles do not overlap, as shown in Fig. 3. The result of the TWIST experiment [12] does not contradict the result of our analysis of neutron decay, since the TWIST experiment was performed with charged particles. The CP-violating effects for $K^0, D^0, B^0, B_s^0$ can be described by the same parameters δ and ζ extracted from neutron decay. They will be discussed below.

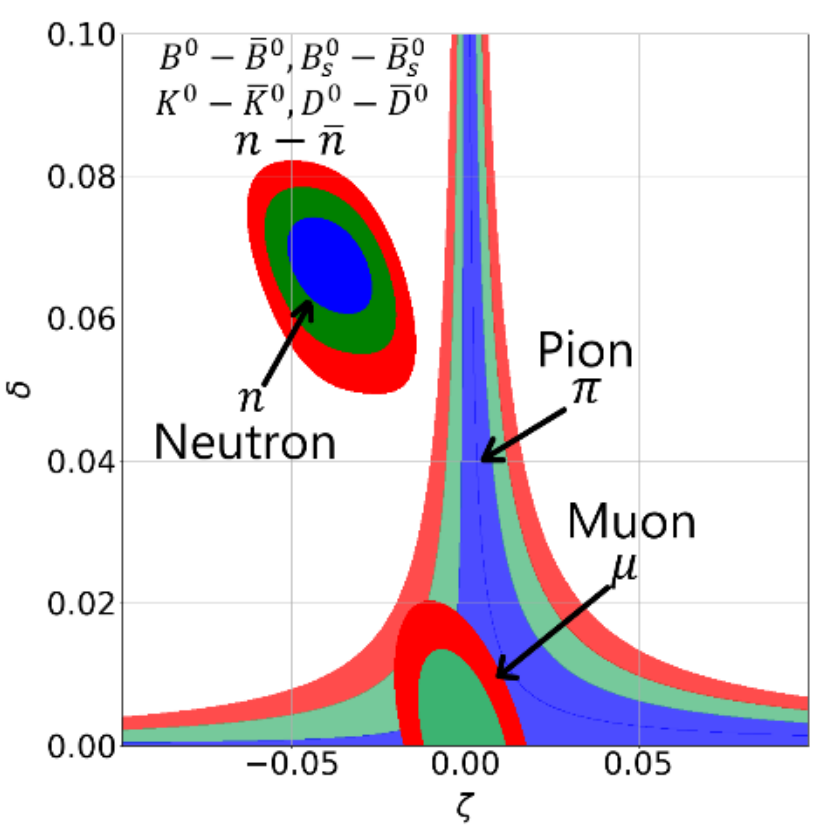

Fig. 3. Comparison of the results of the neutron decay analysis, which are also applicable to the oscillation of the $K^0-\bar{K}^0$, $D^0-\bar{D}^0$, $B^0-\bar{B}^0$, $B_s^0-\bar{B}_s^0$ mesons and also applies to $n-\bar{n}$ oscillations. Besides the constraints from CPT invariance for the $\pi^+\pi^-$ and $\mu^+\mu^-$ decays are shown. The ellipse at the origin is the result of the TWIST experiment in the left-right model interpretation.

The scheme for analyzing neutral meson oscillations in accelerator experiments is represented by matrix (2.13). This is the scheme of the Standard Model. The scheme for analyzing neutral meson oscillations within the extended left-right model is represented by matrix (2.14).

$$\begin{pmatrix} M - i\dfrac{\Gamma_0}{2} & \Delta m - i\dfrac{\Delta\Gamma}{2} \\ \Delta m - i\dfrac{\Delta\Gamma}{2} & M - i\dfrac{\Gamma_0}{2} \end{pmatrix} \quad (2.13)$$

$$\begin{pmatrix} M - i\dfrac{\Gamma_0 - \delta\Gamma}{2} & \Delta m - i\dfrac{\Delta\Gamma}{2} \\ \Delta m - i\dfrac{\Delta\Gamma}{2} & M - i\dfrac{\Gamma_0 + \delta\Gamma}{2} \end{pmatrix} \quad (2.\ 14)$$

The mixing matrix for neutral mesons in the Standard Model takes into account flavor oscillation effects and CP- violation effects in off-diagonal matrix elements, with CP-violation effects being accounted for by the complex phase in the CKM matrix as a phenomenological parameter. The physical causes of CP-violation are unknown. With such a CM matrix, flavor oscillations and CP-violation effects in off-diagonal matrix elements interfere, making it significantly difficult to separate them.

The mixing matrix for neutral mesons in the extended left-right model is structured differently. It takes into account flavor oscillation effects in off-diagonal matrix elements and CP violation effects in diagonal matrix elements through the parameter $\delta\Gamma$.

We believe that the physical cause for $\delta\Gamma$ is related to the different signs of the mixing angle of the right-handed vector boson with the positive and negative left-handed vector bosons, since $A_T^{LR} = -2\delta\zeta = 5.6 \cdot 10^{-3}$, based on the results of the neutron experiment for left-right asymmetry. However, since the experiment [9] demonstrated the absence of CPT violation to an accuracy of $10^{-3}$, we can assume that an effect exceeding this value is primarily due to CP-violation.

Let's compare two schemes for describing the process of neutral meson oscillations. The effective Hamiltonian is written as follows:

$$H = \begin{pmatrix} m_0 - i\dfrac{\Gamma_0}{2} & \Delta m - i\dfrac{\Delta\Gamma}{2} \\ \Delta m - i\dfrac{\Delta\Gamma}{2} & m_0 - i\dfrac{\Gamma_0}{2} \end{pmatrix} \quad (2.15)$$

We will seek a solution to the Schrödinger equation for the Hamiltonian $H\Psi(t) = i\hbar\frac{\partial\Psi(t)}{\partial t}$ in the form of a two-dimensional vector $\Psi(t) = \begin{pmatrix} a \\ b \end{pmatrix} e^{-i\omega t}$. The solution to the Schrödinger equation is reduced to solving the problem of eigenvalues - ω and eigenvectors - $\begin{pmatrix} a \\ b \end{pmatrix}$. The equation for the eigenvalues - ω is determined from the following condition:

$$\begin{pmatrix} m_0 - i\frac{\Gamma_0}{2} & \Delta m - i\frac{\Delta\Gamma}{2} \\ \Delta m - i\frac{\Delta\Gamma}{2} & m_0 - i\frac{\Gamma_0}{2} \end{pmatrix} \begin{pmatrix} a \\ b \end{pmatrix} = \omega \begin{pmatrix} a \\ b \end{pmatrix} \quad (2.16),$$

$$\det \begin{Vmatrix} m_0 - i\dfrac{\Gamma_0}{2} - \omega & \Delta m - i\dfrac{\Delta\Gamma}{2} \\ \Delta m - i\dfrac{\Delta\Gamma}{2} & m_0 - i\dfrac{\Gamma_0}{2} - \omega \end{Vmatrix} = 0 \quad (2.17),$$

Eigenvalues: $\omega_\pm = m_0 \pm \Delta m - i\frac{\Gamma_0 \pm \Delta\Gamma}{2}$.

Frequency difference: $\omega_+ - \omega_- = \Delta\omega = 2\Delta m - i\Delta\Gamma$ .

For $\omega_+ = m_0 + \Delta m - i\frac{\Gamma_0 + \Delta\Gamma}{2}$ the normalized eigenvector is equal to:

$$\Psi_L(t) = \frac{1}{\sqrt{2}}\begin{pmatrix} 1 \\ 1 \end{pmatrix} e^{-i\,\omega_+ t}$$

For $\omega_- = m_0 - \Delta m - i\frac{\Gamma_0 - \Delta\Gamma}{2}$ the normalized eigenvector is equal to:

$$\Psi_S(t) = \frac{1}{\sqrt{2}}\begin{pmatrix} 1 \\ -1 \end{pmatrix} e^{-i\omega_- t}$$

If at time t=0 we have the state of a particle, then

$$\psi(t) = \frac{1}{\sqrt{2}}\left(\Psi_L(t) + \Psi_S(t)\right) \quad (2.18).$$

If at the initial moment of time the state of the antiparticle is, then

$$\bar{\psi}(t) = \frac{1}{\sqrt{2}}\left(\Psi_L(t) - \Psi_S(t)\right) \quad (2.19).$$

When the particle is in the initial state, the probability of detecting the particle will be (2.20), and the probability of detecting the antiparticle will be equal to (2.21).

$$\psi^*(t)\psi(t) = \frac{1}{4}e^{-\Gamma_0 t}[e^{-\Delta\Gamma t} + e^{\Delta\Gamma t} + 2\cos(2\Delta m t)] \quad (2.20),$$

$$\bar{\psi}^*(t)\bar{\psi}(t) = \frac{1}{4}e^{-\Gamma_0 t}[e^{-\Delta\Gamma t} + e^{\Delta\Gamma t} - 2\cos(2\Delta m t)] \quad (2.21),$$

When the initial state is an antiparticle, the probability of detecting the antiparticle will be (2.22), and the probability of detecting the particle will be (2.23).

$$\bar{\psi}^*(t)\bar{\psi}(t) = \frac{1}{4}e^{-\Gamma_0 t}[e^{-\Delta\Gamma t} + e^{\Delta\Gamma t} + 2\cos(2\Delta m t)] \quad (2.22),$$

$$\psi^*(t)\psi(t) = \frac{1}{4}e^{-\Gamma_0 t}[e^{-\Delta\Gamma t} + e^{\Delta\Gamma t} - 2\cos(2\Delta m t)] \quad (2.23)$$

It is clear that particles and antiparticles behave symmetrically: when the initial conditions change, it is equivalent the particle and antiparticle swap places. We will use this fact to estimate the magnitude of the asymmetry between the decay probabilities of particles and antiparticles, taking mixing into account.

The next step is to introduce CP-violation. In the Standard Model, CP-violation is introduced through the complex phase in the CKM matrix. However, we propose introducing CP-violation into the diagonal elements of the mixing matrix. Regarding the CKM matrix, we would expect a splitting of the matrix elements for interactions via $W^+$ and $W^-$. Otherwise, appropriate corrections would need to be made.

Let's analyze the results of experiments with neutral mesons using the mixing matrix representation (2.14), in which the parameters responsible for flavor oscillations (ΔΓ and Δm) and the parameters responsible for CP-violation effects (δΓ) are separated. Moreover, the parameters ΔΓ and Δm are taken from experimental data, and the parameter $\delta\Gamma = -2\delta\zeta \times \Gamma = 5.5 \times 10^{-3} \times \Gamma$. It is important to emphasize that it is here that the parameters of the extended left-right model are used, and in a unified manner to describe the CP-violation effect in the oscillations of all neutral mesons.

Thus, if we introduce δΓ for the particle and antiparticle in the diagonal matrix elements of the Hamiltonian, we can write:

$$H = \begin{pmatrix} m_0 - i\frac{\Gamma_0 - \delta\Gamma}{2} & \Delta m - i\frac{\Delta\Gamma}{2} \\ \Delta m - i\frac{\Delta\Gamma}{2} & m_0 - i\frac{\Gamma_0 + \delta\Gamma}{2} \end{pmatrix} \quad (2.24).$$

The different signs of δΓ mean that the sign of the CP-violation effect changes when moving from particle to antiparticle. Thus, in the presence of sufficiently frequent oscillations, the average lifetime for particles and antiparticles is the same, and the CP-violation effect averages out to zero. However, when the lifetime is significantly shorter than the oscillation period, CP-violation occurs.

Thus, the criterion for the validity of the CP-violation scheme within the extended left-right model is as follows:

No. 1. The integral (average value over several decay periods) CP-violation effect is practically zero when the oscillation period is significantly shorter than the decay time.

No. 2. The integral CP-violation effect manifests itself when the oscillation period is significantly longer than the decay time.

No.3. The differential CP-violation effect (dependence on the decay time) reverses its sign upon transition from particle to antiparticle, when the oscillation period is significantly shorter than the decay time.

These conclusions follow from the fact that the transition from particle to antiparticle occurs through a positive or negative vector boson, each with a different sign of the mixing angle ζ with the mixing in right-handed vector boson, which changes the sign of the CP-violation effect.

Our task is to calculate the behavior of the integral and differential CP-violation effects depending on the above conditions and compare them with experimental results.

We now turn to solving the problem of CP-violating oscillations within the extended left-right model. Modifying the Hamiltonian with the δΓ corrections will change both the eigenvalues and eigenvectors in the previous problem:

$$\omega_+ = m_0 - i\frac{\Gamma_0}{2} + \frac{\Delta\omega}{2},$$
$$\omega_- = m_0 - i\frac{\Gamma_0}{2} - \frac{\Delta\omega}{2},$$

where

$$\Delta\omega = \sqrt{(2\Delta m)^2 - [(\Delta\Gamma)^2 + (\delta\Gamma)^2] - 2i(\Delta\Gamma)(2\Delta m)} \quad (2.25)$$

The corrections from δΓ leads to the fact that in the first order in δΓ the probability of detecting a particle, calculated in the absence of δΓ in the Hamiltonian, receives an addition $\varepsilon_{pp}(t)$

$$\psi^*(t)\psi(t) \approx \frac{1}{4}e^{-\Gamma_0 t}\left[e^{\Delta\Gamma t} + e^{-\Delta\Gamma t} + 2cos(2\Delta m t)\right] + \varepsilon_{pp}(t) \quad (2.26)$$

For an antiparticle, the probability taking into account δΓ will have a contribution $\varepsilon_{pp}(t)$ with a different sign.

$$\bar{\psi}^*(t)\bar{\psi}(t) \approx \frac{1}{4}e^{-\Gamma_0 t}\left[e^{-\Delta\Gamma t} + e^{\Delta\Gamma t} + 2\cos(2\Delta m t)\right] - \varepsilon_{pp}(t) \quad (2.27)$$

For approximate calculations of asymmetry in the expansion in terms of the small parameter

$x = \delta\Gamma/((2\Delta m)^2 + (\Delta\Gamma)^2)^{1/2}$ we use the formula:

$$\varepsilon_{pp}(t) = \frac{\delta\Gamma}{\sqrt{(2\Delta m)^2+(\Delta\Gamma)^2}} \times \frac{e^{-\Gamma_0 t}\left((\Delta\Gamma)sh(\Delta\Gamma t)+(2\Delta m)sin(2\Delta m t)\right)}{\sqrt{(2\Delta m)^2+(\Delta\Gamma)^2}} \quad (2.28)$$

The results of calculations for the probability of detecting a particle or antiparticle for the two schemes are presented in Fig. 4. As can be seen, the results of calculations with $\delta\Gamma$=0 and $\delta\Gamma/\Gamma = -2\delta\zeta = 5.5 \times 10^{-3}$ do not differ within the limits of graphical accuracy for all mesons. The fact is that the process of flavor oscillations is successfully described due to the off-diagonal parameters of the mixing matrices, and the diagonal parameter $\delta\Gamma$, responsible for CP violation, has little effect on the oscillation process. However, as will be shown below, the diagonal parameter $\delta\Gamma$ plays a decisive role in calculating the CP-violating decay asymmetry. Thus, in the mixing matrix (2.14) in the extended left-right model, flavor oscillations and CP violation effects are separated to a certain extent.

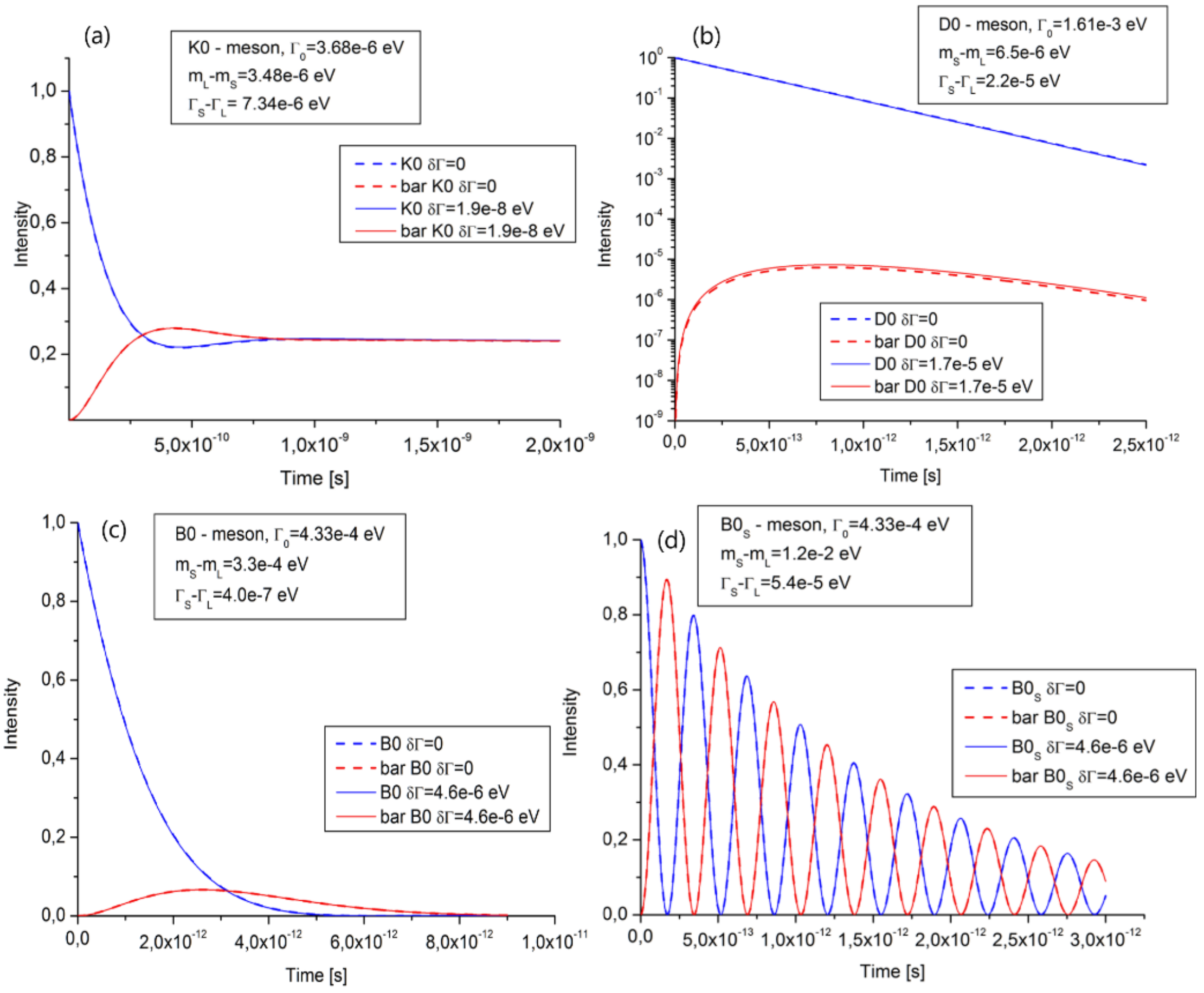

Fig. 4. The process of neutral meson oscillations. Blue curves correspond to the particle state, red curves – to the antiparticle state, solid curves correspond to the calculations carried out with the matrix (2.13) (CM), dashed curves correspond to the calculations carried out with the matrix (2.14) (LRM). In the calculations, the off-diagonal matrix elements are defined with $2\Delta m = (m_S - m_L)$, $2\,\Delta\Gamma = (\Gamma_S - \Gamma_L)$, a) – the calculation results for the $K^0$-meson, b) – for the $D^0$-meson, c) – for the $B^0$-meson, d) - for the $B_S^0$-meson.

## 3. Behavior of integral and differential effects of CP violation and comparison with experimental results

Now let's move on to calculating the integral asymmetry. For this, we take the ratio of the integral $\varepsilon_{pp}$ to the integral $\rho_{pp}$.

$$A_{pp\ \bar{p}\bar{p}} = \frac{\left|\psi_{up,p}\right|^2 - \left|\tilde{\psi}_{up,\bar{p}}\right|^2}{\left|\psi_{up,p}\right|^2 + \left|\tilde{\psi}_{up,\bar{p}}\right|^2},\ \tilde{A}_{p\bar{p}} = \frac{\varepsilon_{pp}}{\rho_{pp}},$$

where

$$\varepsilon_{pp} = \int_0^\infty \varepsilon_{pp}(t)dt,\ \rho_{pp} = \int_0^\infty |\psi(t)|^2 dt \quad (3.1)$$

Table 1 lists the integral asymmetries $\tilde{A}_{p\bar{p}}$ for the $K^0$, $D^0$, $B^0$ and $B_S^0$ mesons, calculated using the wave functions (Exact) and the approximate formulas (Approx). As can be seen, the approximate formulas agree reasonably well with the exact computer calculations. However, it is important to note that the formulas allow for an analytical consideration of various dependencies and relations, which are extremely useful.

The experiment measures the effects of CP-violation as a function of the decay time, so it is necessary to calculate these effects as a function of time. To compare the differential effects of CP-violation with the experimental results, the dependence of the asymmetry on the decay time was calculated. The dependence of the differential asymmetry on the decay time is represented by expression (3.2), obtained from equations (2.26), (2.27), and (2.28).

$$f_{asym}(t) = \frac{2\delta\Gamma}{\sqrt{(\Delta\Gamma)^2+(2\Delta m)^2}} \times \frac{(\Delta\Gamma)\mathrm{sh}(\Delta\Gamma t)+(2\Delta m)\sin(2\Delta m t)}{\sqrt{(\Delta\Gamma)^2+(2\Delta m)^2}[\mathrm{ch}(\Delta\Gamma t)+\cos(2\Delta m t)]} \qquad (3.2)$$

The absolute values of calculated and experimental asymmetries are quite difficult to compare. The experiment uses a specific decay mode, while the calculation considers particle-antiparticle transitions, which cover all decay channels. Therefore, there is a problem with normalizing the asymmetry value. For example, for the $B_S^0$ meson the calculated asymmetry will have a resonant nature during the particle-antiparticle transition, since the particle probability at this point drops to zero. However, the experimental asymmetry is normalized to the decay probability. When normalized to the decay probability, the calculated asymmetry will have the form (3.2), but $\cos(2\Delta mt) = 1$. Therefore, to correctly compare calculations with experimental data, the following formula should be used:

$$\begin{aligned} A_{CP}(t) &= \frac{2\delta\Gamma}{\sqrt{(\Delta\Gamma)^2+(2\Delta m)^2}} \times \frac{(\Delta\Gamma)\mathrm{sh}(\Delta\Gamma t)+(2\Delta m)\sin(2\Delta m t)}{\sqrt{(\Delta\Gamma)^2+(2\Delta m)^2}[\mathrm{ch}(\Delta\Gamma t)+1]} = \\ &= (-2\delta\zeta)\frac{2\Gamma}{\sqrt{(\Delta\Gamma)^2+(2\Delta m)^2}} \times \frac{(\Delta\Gamma)\mathrm{sh}(\Delta\Gamma t)+(2\Delta m)\sin(2\Delta m t)}{\sqrt{(\Delta\Gamma)^2+(2\Delta m)^2}[\mathrm{ch}(\Delta\Gamma t)+1]} = \\ &= A^{LR} \times F \times A_{asym}(t) \end{aligned} \qquad (3.3)$$

where $A^{LR} = -2\delta\zeta$ there is a CP-violating asymmetry of the left-right weak interaction model

$F = \frac{2\Gamma}{\sqrt{(\Delta\Gamma)^2+(2\Delta m)^2}}$ - a time-independent factor characterizing a specific meson (3.4),

$A_{asym}(t) = \frac{(\Delta\Gamma)\mathrm{sh}(\Delta\Gamma t)+(2\Delta m)\sin(2\Delta m t)}{\sqrt{(\Delta\Gamma)^2+(2\Delta m)^2}[\mathrm{ch}(\Delta\Gamma t)+1]}$ - time-dependent asymmetry characterizing a specific meson (3.5).

The integral value of the CP-violating asymmetry is: $\mathrm{A_{CP}} = \int A_{CP}(t)\mathrm{e}^{-t/\tau}dt$ (3.6)

Table 1 presents the time-independent factor $F$ and the integral value of the CP-violating asymmetry $\mathrm{A_{CP}}$. Time-dependent asymmetry functions $\mathrm{A_{CP}}(t)$ for all mesons will be presented later.

Table 1. $(2\Delta m = m_S - m_L, 2\Delta\Gamma = \Gamma_S - \Gamma_L)$

| Meson | $2\Delta m[eV]$ | $2\Delta\Gamma[eV]$ | $\Gamma$ , $\delta\Gamma[eV]$ | Exact $\tilde{A}_{p\bar{p}}$ | Approx $\tilde{A}_{p\bar{p}}$ | $\Gamma / 2\Delta m = T_{осц} / \tau_{расп}$ |
|---|---|---|---|---|---|---|
| $K^0$ | $3.48 \times 10^{-6}$ | $7.3 \times 10^{-6}$ | $3.68 \times 10^{-6}$ $(1.9 \times 10^{-8})$ | $5.5 \times 10^{-3}$ | $5.6 \times 10^{-3}$ | 2.13 |
| $D^0$ | $6.5 \times 10^{-6}$ | $2.2 \times 10^{-5}$ | $1.61 \times 10^{-3}$ $(8.5 \times 10^{-6})$ | $5.3 \times 10^{-3}$ | $5.3 \times 10^{-3}$ | 250 |
| $B^0$ | $3.3 \times 10^{-4}$ | $4.0 \times 10^{-7}$ | $4.33 \times 10^{-4}$ $(2.3 \times 10^{-6})$ | $3.3 \times 10^{-3}$ | $3.4 \times 10^{-3}$ | 1.32 |
| $B_S^0$ | $1.2 \times 10^{-2}$ | $5.4 \times 10^{-5}$ | $4.33 \times 10^{-4}$ $(2.3 \times 10^{-6})$ | $\mathbf{6.9 \times 10^{-6}}$ | $\mathbf{6.9 \times 10^{-6}}$ | 0.036 |

These calculations used the PDG values [13] presented in Tables 2 and 3.

Table 2

| | Values in calculations | | Data from PDG | |
|---|---|---|---|---|
| Мезон | $m[eV]$ | $2\Delta m[eV]$ | $m$ | $2\Delta m$ |
| $K^0$ | $497.611 \times 10^6$ | $3.48 \times 10^{-6}$ | **$K^0$ MASS** VALUE (MeV) **497.611±0.013** | $m_{K_L^0} - m_{K_S^0}$ VALUE ($10^{10}$ $\hbar$ $s^{-1}$) **0.5293 ±0.0009** $\boldsymbol{\Delta m = (3.48 \pm 0.01) \times 10^{-6} eV}$ |
| $D^0$ | $1864.84 \times 10^6$ | $6.5 \times 10^{-6}$ | **$D^0$ MASS** VALUE (MeV) **1864.84 ±0.05** | $\lvert m_{D_1^0} - m_{D_2^0} \rvert$ VALUE ($10^{10}$ $\hbar$ $s^{-1}$) **0.997±0.116** $\boldsymbol{\Delta m = (6.5 \pm 0.8) \times 10^{-6} eV}$ |
| $B^0$ | $5279 \times 10^6$ | $3.3 \times 10^{-4}$ | **$B^0$ MASS** VALUE (MeV) **5279.72±0.08** | $\Delta m_{B^0} = m_{B_H^0} - m_{B_L^0}$ VALUE ($10^{12}$ $\hbar$ $s^{-1}$) **0.5069±0.0019** $\boldsymbol{\Delta m = (3.34 \pm 0.01) \times 10^{-4} eV}$ |
| $B_S^0$ | $5366 \times 10^6$ | $1.2 \times 10^{-2}$ | **$B_s^0$ MASS** VALUE (MeV) **5366.93± 0.10** | $\Delta m_{B_s^0} = m_{B_{sH}^0} - m_{B_{sL}^0}$ VALUE ($10^{12}$ $\hbar$ $s^{-1}$) **17.765 ±0.006** $\boldsymbol{\Delta m = (1.17 \pm 0.04) \times 10^{-2} eV}$ |

Table 3

<table>
<tr><td></td><td colspan="2">Values in calculations</td><td colspan="3">Data from PDG</td></tr>
<tr><td>Мезон</td><td>$\Gamma[eV]$</td><td>$2\Delta\Gamma[eV]$</td><td colspan="2">$\Gamma$</td><td>$2\Delta\Gamma$</td></tr>
<tr><td>$K_S^0$</td><td>$7.35 \times 10^{-6}$</td><td rowspan="2">$7.35 \times 10^{-6}$</td><td rowspan="2">$K_L^0$ MEAN LIFE<br>VALUE ($10^{-8}$ s)<br>5.116±0.021<br>$\boldsymbol{\Gamma = (1.29 \pm 0.03) \times 10^{-8} eV}$</td><td rowspan="2">$K_S^0$ MEAN LIFE<br>VALUE ($10^{-10}$ s)<br>0.8954 ±0.0004<br>$\boldsymbol{\Gamma = (7.35 \pm 0.01) \times 10^{-6} eV}$</td><td rowspan="2">$\boldsymbol{\Gamma_S \gg \Gamma_L \to \Delta\Gamma \approx \Gamma_S}$<br>$\boldsymbol{\Gamma = (7.35 \pm 0.01) \times 10^{-6}\ eV}$</td></tr>
<tr><td>$K_L^0$</td><td>$1.29 \times 10^{-8}$</td></tr>
<tr><td>$D^0$</td><td>$1.60 \times 10^{-6}$</td><td>$2.2 \times 10^{-5}$</td><td colspan="2">$D^0$ MEAN LIFE VALUE ($10^{-15}$ s) 410.3± 1.0<br>$\boldsymbol{\Gamma = (1.60 \pm 0.01) \times 10^{-6} eV}$</td><td>$(\Gamma_{D_1^0} - \Gamma_{D_2^0})/\Gamma$ VALUE (units $10^{-2}$) 1.394± 0.056<br>$\boldsymbol{\Delta\Gamma = (2.2 \pm 0.01) \times 10^{-5}\ eV}$</td></tr>
<tr><td>$B^0$</td><td>$4.34 \times 10^{-4}$</td><td>$4.0 \times 10^{-7}$</td><td colspan="2">$B^0$ MEAN LIFE VALUE ($10^{-12}$ s) 1.517±0.004<br>$\boldsymbol{\Gamma = (4.34 \pm 0.01) \times 10^{-4} eV}$</td><td>$\Delta\Gamma_{B_d^0} / \Gamma_{B_d^0}$ VALUE (units $10^{-2}$) 0.1 ±1.0<br>$\boldsymbol{\Delta\Delta\Gamma = (4 \pm 40) \times 10^{-7}\ eV}$</td></tr>
<tr><td>$B_S^0$</td><td>$4.34 \times 10^{-4}$</td><td>$5.4 \times 10^{-5}$</td><td colspan="2">$B_s^0$ MEAN LIFE VALUE ($10^{-12}$ s) 1.516±0.006<br>$\boldsymbol{\Gamma = (4.34 \pm 0.01) \times 10^{-4} eV}$</td><td>$\Delta\Gamma_{B_s^0}/\Gamma_{B_s^0}$ VALUE 0.124±0.007<br>$\boldsymbol{\Delta\Gamma = (5.4 \pm 0.3) \times 10^{-5}\ eV}$</td></tr>
</table>

Let's discuss the results in Table 1. We can see that for the $K^0$-meson, $D^0$– meson, and $B^0$– meson, we have integral CP asymmetries that are practically determined by the asymmetry value: $A_T^{LR} = -2\delta\zeta = 5.5 \times 10^{-3}$. However, for the $B_S^0$– meson, the CP-violation effect is $6.9 \times 10^{-6}$, since the CP asymmetry averages out during the oscillations of the $B_S^0$–meson. Nevertheless, it should be noted that the CP violation process existed throughout the entire oscillation process. The fact is that the sign of the CP-violating effect changed to the opposite during each oscillation period, in accordance with the formula $\varepsilon_{p\bar{p}} \approx e^{-\Gamma_0 t} \sin(2\Delta m t) \left(\frac{\delta\Gamma}{2\Delta m}\right)$, and therefore was compensated. Experimental observations with more data in 2021 confirmed the absence of asymmetry [11] with an accuracy of $1.1 \times 10^{-3}$ (Fig. 5).

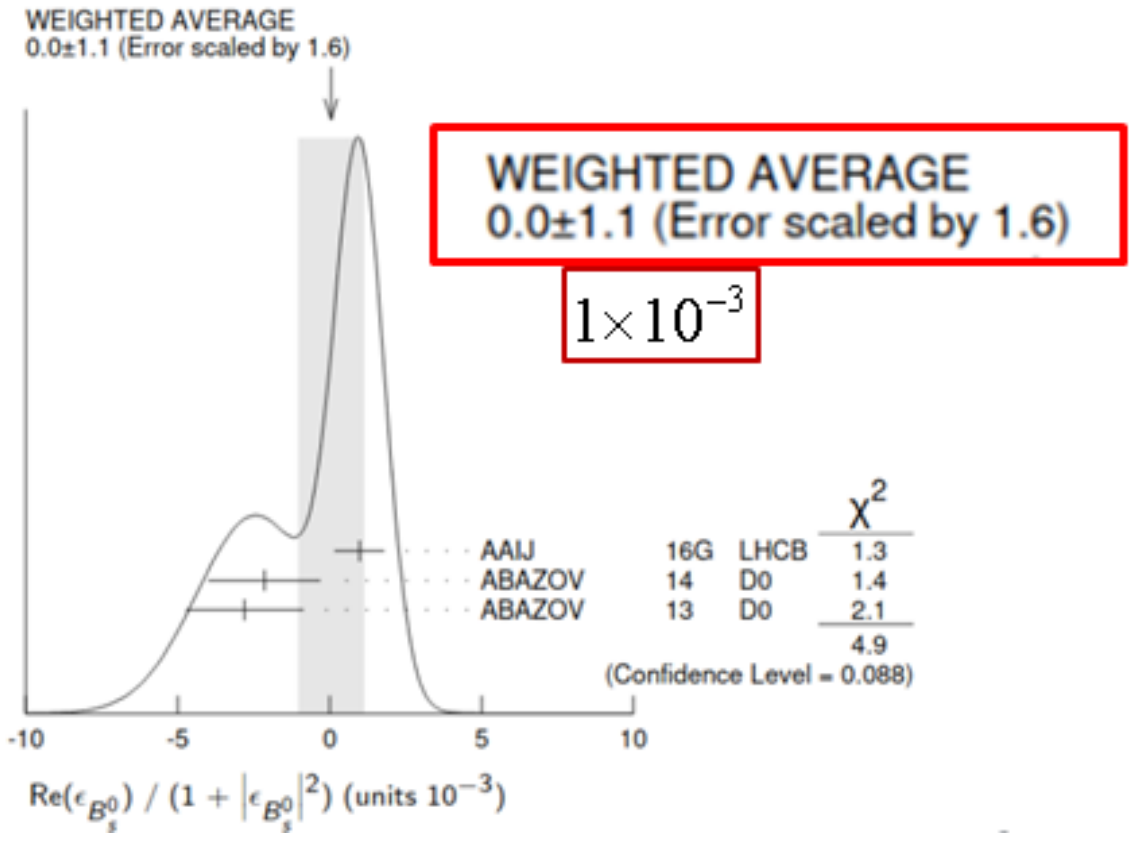

Fig. 5. Experimental data from the PDG for the integral CP violation effect for the $B_S^0$ meson.

For the $K^0$meson, the integral asymmetry calculation yielded a result of $A_{CP}^{K^0}$= $5.6 \times 10^{-3}$. The calculation of integral asymmetries is simplified to a certain extent. The reason is that for the $K^0$ meson the CP-violation effect can be represented as a mixture of two effects: a mixing effect for the $K_S^0$-meson, when the lifetime is much shorter than the oscillation period of the $K_L^0$-meson, and a final state effect for the $K_L^0$-meson, when the lifetime practically coincides with the oscillation period. Indeed, there is an experimental result for mixing: $A_T^{exp} = (6.6 \pm 1.3 \pm 1.0) \times 10^{-3}$ [14], in addition, there is an experimental result for the decay in the final state: $A_{\text{CP}}^{exp} = (3.32 \pm 0.06) \times 10^{-3}$ [13]. For the $K_L^0$-meson, the effect becomes smaller due to the manifestation of the oscillation process.

Thus, for the $K^0$-meson, criterion No.2 is fulfilled (The integral effect of CP violation appears when the oscillation period is significantly greater than the decay time). In this case, $\Gamma/2\Delta m = T_{osc}/\tau_{decay} = 2.13$.

In addition, for the $K^0$ meson, in the same work [7] there is an experimental dependence of the CP-violating asymmetry effect on the decay time - $A_{CP}(t)$ . It is shown in Fig. 6 in the interval of 20 lifetimes of the $K_S$ meson. The average value of the asymmetry is $(6.6 \pm 1.3 \pm 1.0) \times 10^{-3}$ [7]. Our calculated dependence of the CP-violating asymmetry effect on the decay time - $A_{CP}(t)$ is shown in the same figure by the blue line.

It can be concluded that the experimental results for the $K^0$meson, within the available accuracy, can be described in the left-right weak interaction model.

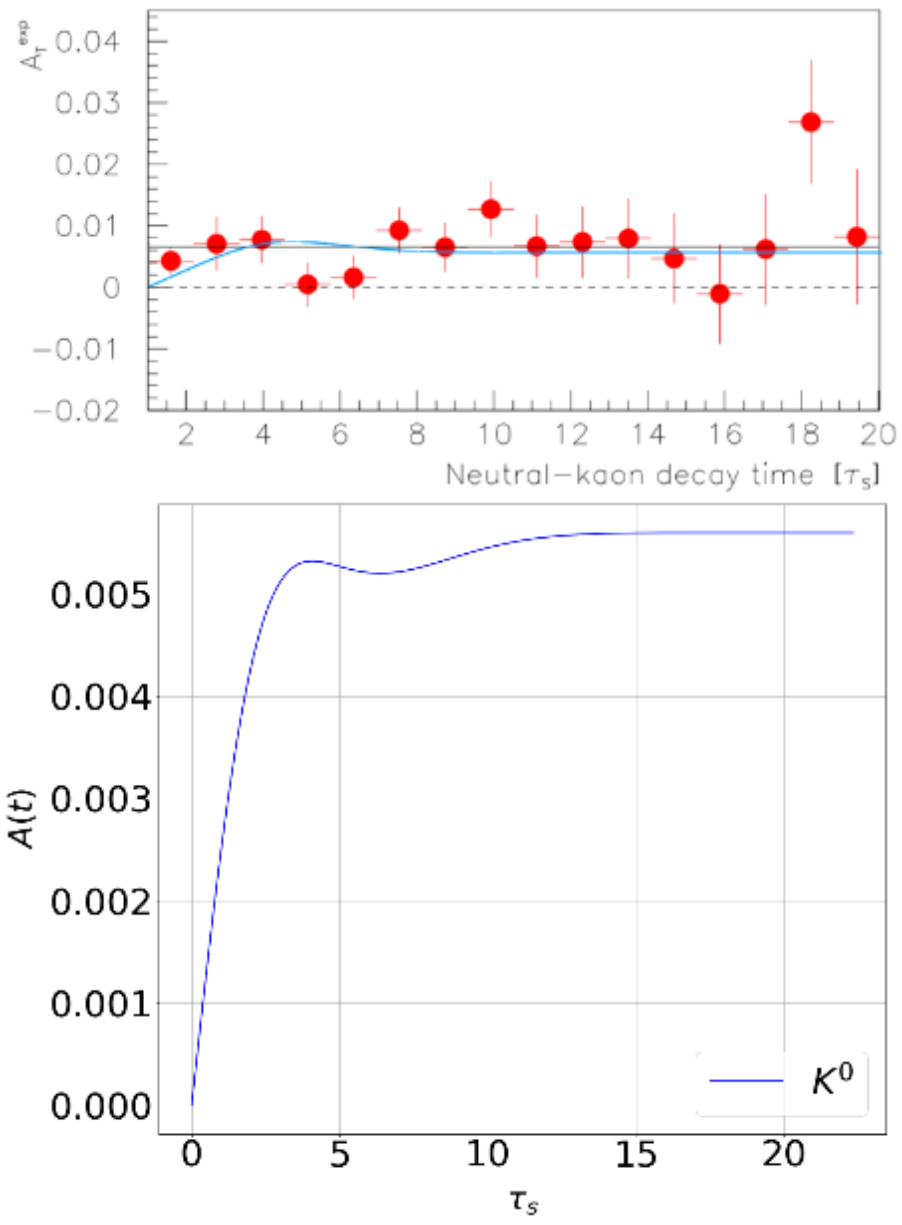

Fig. 6. Calculated dependence of the CP-violating asymmetry effect on the decay time for the $K^0$meson.

For the $D^0$meson, the ratio of the decay rate to the oscillation frequency is $\Gamma / 2\Delta m = T_{осц} / \tau_{расп}$, so the oscillation process has time to decay before it even begins. In this case, criterion No.2 is satisfied with good accuracy for the $D^0$ meson (The integral CP-violation effect manifests itself when the oscillation period is significantly longer than the decay time). The experimental value of the CP-violation effect is A=(6.8±0.65) $10^{-3}$ [15], while the calculated value is $A_{CP}^{D^0} = 5.3 \times 10^{-3}$. The calculated dependence of the CP violation effect on the decay time, which was used for integration, is shown in Fig. 7. As can be seen, the asymmetry at large times ($2 \cdot 10^{-10}$ s) reaches 68%, but the decay time of the $D^0$ meson is $\tau(D^0) = 4.1 \times 10^{-13} s$ , therefore, only the initial portion of the differential asymmetry at contributes to the integral value of the asymmetry. Thus, $A_{CP}^{D^0} = \int A_{CP}(t) e^{-t/\tau} dt$ =5.3 × $10^{-3}$. It can be concluded that the experimental results for the $D^0$ meson within the available accuracy, can be described by the left-right weak interaction model.

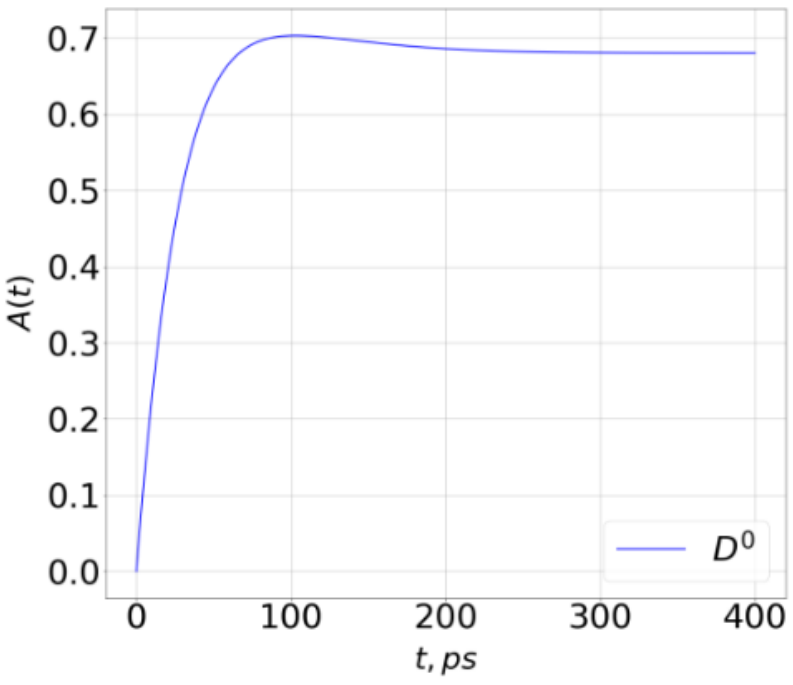

Fig. 7. Calculated dependence of the CP violation effect on the decay time for the $D^0$meson, which was used for integration.

For the $B^0$meson, the ratio of the decay rate to the oscillation frequency is 1.32, so the oscillation process is on the verge of satisfying criterion No.2 (the integral effect of CP-violation manifests itself when the oscillation period is significantly longer than the decay time), while criterion No.3 is quite clearly satisfied (when the differential effect of CP-violation changes the sign of the effect to the opposite upon the transition from particle to antiparticle). The experimental effect of CP-violation as a function of the decay time is shown in Fig. 8, and the calculated time-dependent component of the asymmetry:

$$A_{asym}(t) = \frac{(\Delta\Gamma)\,\mathrm{sh}(\Delta\Gamma t) + (2\Delta m)\sin(2\Delta m t)}{\sqrt{(\Delta\Gamma)^2 + (2\Delta m)^2}\,[\mathrm{ch}(\Delta\Gamma t) + 1]},$$

characterizing the $B^0$meson is shown in the Fig. 8.

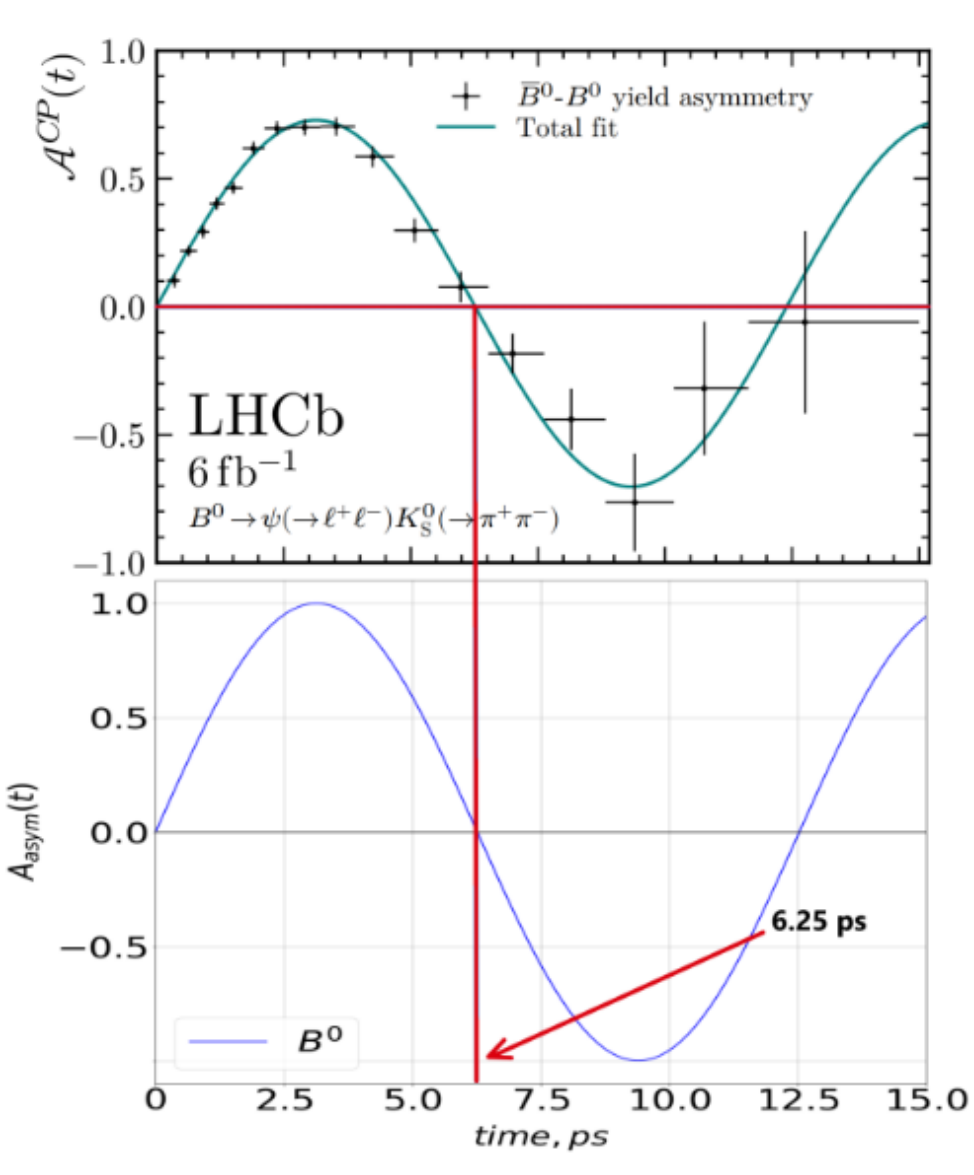

Fig. 8. Comparison of the measured time dependence of the $B^0 \to \psi K_S^0$ decay asymmetry from [16] and the prediction of the time-dependent component of the asymmetry $A_{asym}(t)$ for the $B^0$ meson within the left-right model.

As can be seen from Fig. 8, the experimental time-dependent asymmetry reaches a fairly large level - much larger than the average value of the CP-violating asymmetry $A^{LR}$ over all decay modes in the $B^0 - \bar{B}^0$oscillation process. However, it should be noted that the experimental result is presented for a decay mode that accounts for only 5 · $10^{-5}$ of the total decay probability of the $B^0$meson. Therefore, the contribution to the average integral value of the CP-violating asymmetry over all decay modes in the $B^0 - \bar{B}^0$ oscillation process will be significantly suppressed. We also note that the average integral value of the CP-violating asymmetry for the $B^0$meson in the PDG data is 0.005±0.012±0.014 [13].

Thus, for the $B^0$ meson, we have fairly clear experimental confirmation that criterion #3 is satisfied, i.e. change in the sign of CP violation occurs upon the transition from $B^0$ to $\bar{B}^0$ .

Finally, for the $B_s^0$ meson, the ratio of the decay rate to the oscillation frequency is $\Gamma / 2\Delta m = T_{осц} / \tau_{расп}$ = $3.6 \times 10^{-2}$, i.e. much less than unity, and therefore the oscillation process is represented by a large number of periods. However, due to the high oscillation frequency, it is difficult to obtain good experimental statistical significance as for the $B^0$meson. A clear synchronization in the change of sign of CP-violation is demonstrated in the calculations for the $B_s^0$-meson in Fig. 9.

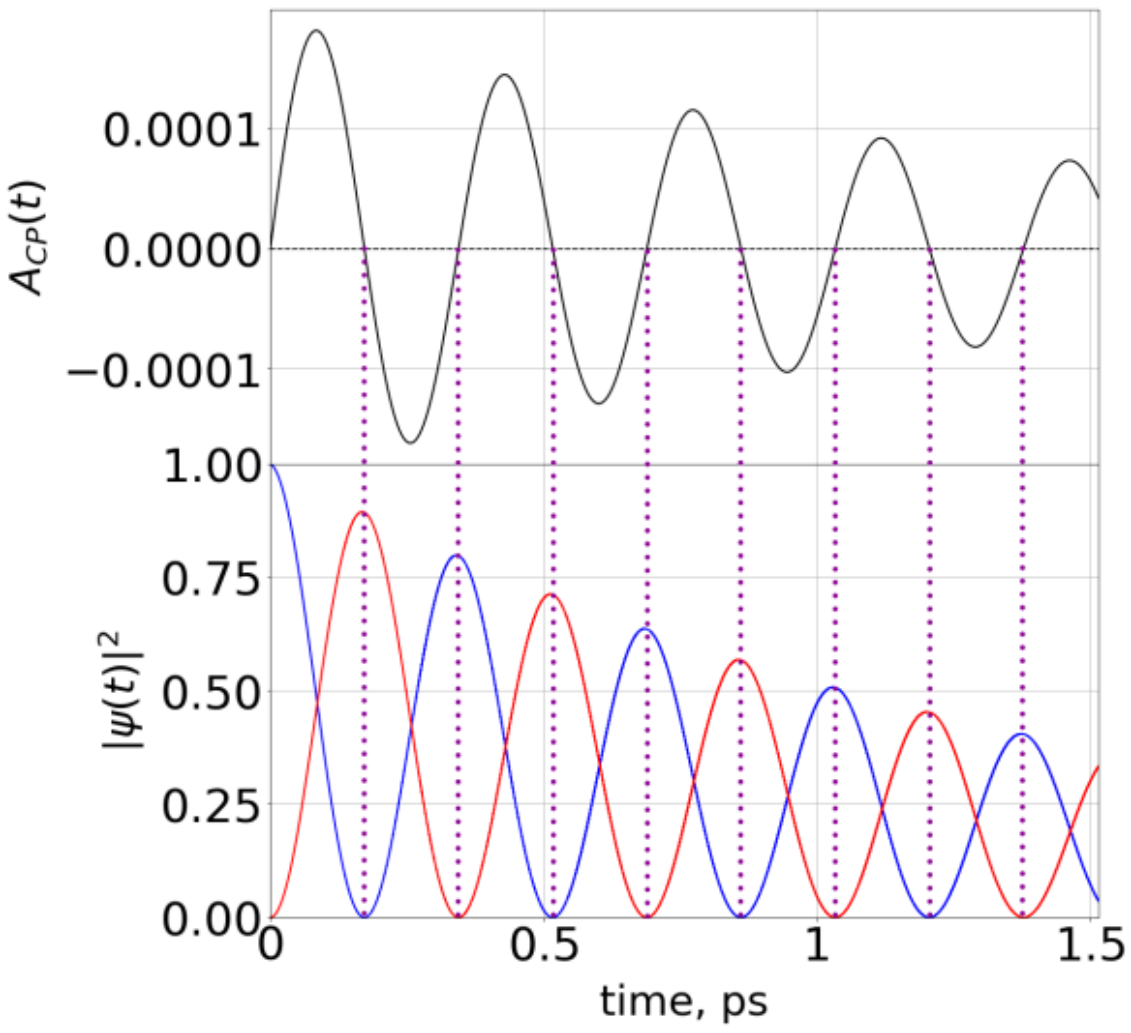

Fig. 9. CP-violating asymmetry for the $B_s^0$ meson and the probabilities of states during oscillations for a particle (blue curve) and an antiparticle (red curve). A clear synchronization in the change in the sign of CP violation for the $B_s^0$ meson is observed: the positive sign of CP violation changes to a negative sign at the time when the transition process from particles to antiparticles changes to a transition process from antiparticles to particles (shown by the dotted line).

As can be seen, the positive sign of CP-violation changes to a negative sign at the time 0.17 ps, when the transition process from particles to antiparticles changes to a transition process from antiparticles to particles. The new sign of the CP-violation process is determined by the direction of the transition: from plus to minus during the transition from particles to antiparticles (at 0.17 ps) and from minus to plus during the transition from antiparticles to particles (at 0.17 ps). This is the essence of the CP-violation process, since the sign of CP-violation depends on the direction of time during which the process occurs, i.e., forward or backward relative to the current time.

## 4.Conclusion

Thus, we can conclude that the results of calculations within the extended left-right model are confirmed by experimental results. The extended left-right model with CP-violation assumed a change in the sign of mixing with the right-handed vector boson when moving from $W^-$ (particle) to $W^+$ (antiparticle). This leads to the fact that mixing with quarks via $W^-$ results in CP-violation with a positive sign, while mixing with quarks via $W^+$ results in CP-violation with a negative sign. The nature of CP violation is related to the presence of an admixture of the right-handed vector boson with different signs of the mixing angle.

For comparison with the CP-violation scheme in the Standard Model, it should be noted that CP-violation effects are taken into account through the complex phase in the CKM matrix as a phenomenological parameter, while the physical causes of CP-violation are unknown. Therefore, a further task is to incorporate the extended left-right model into the Standard Model.

## Acknowledgments

The authors thank the staff of the High Energy Physics Division of the Petersburg Nuclear Physics Institute for discussions of this work at the HEPD seminars and, in particular, A.A. Dzyuba for numerous discussions.

## Funding

This work was supported by the Russian Science Foundation (Project No. 24-12-00091 https://rscf.ru/project/24-12-00091/).

## Conflict of Interest

The authors declare no conflicts of interest.

## References

[1] Serebrov A. P., Zherebtsov O. M., Fomin A. K., Samoilov, R. M. Budanov, N. S., Phys. Rev. D 112, 115012

[2] M. A. B. Beg, R. V. Budny, R.N. Mohapatra, and A. Sirlin, Phys. Rev. Lett. 38, 1252 (1977)

[3] B. R. Holstein and S. B. Treiman, Phys. Rev. D 16, 2369 (1977)

[4] P. Herczeg, Phys. Rev. D 34, 3449 (1986)

[5] J. C. Hardy and I. S. Towner, Phys. Rev. C 91, 025501 (2015)

[6] J. C. Hardy and I. S. Towner, Phys. Rev. C 102, 045501 (2020)

[7] A. Angelopoulos et al. Eur. Phys. J. C 22 (2001) 55-79

[8] D.S Ayres, et al, Phys.Rev.D 3, 1051 (1971)
[9] A. Angelopoulos, et al, Phys. Lett. B 471, 332 (1999)
[10] Phys. Lett. B 458 (1999) 545-552
[11] LHCb collaboration, Nature Physics volume 18, pages1–5 (2022)
[12] J. F. Bueno, R. Bayes, Yu. I. Davydov, et al, Phys. Rev. D 84, 032005 (2011)
[13] S. Navas et al. (Particle Data Group), Phys. Rev. D 110, 030001 (2024)
[14] A. Angelopoulos Phys. Lett. B 444 38-42 (1998)
[15] LHCb collaboration, JHEP 12 (2021) 141
[16] R. Aaij, Phys. Rev. Lett. 132 (2024) 021801